\documentclass[preprint]{elsarticle}
\usepackage{lineno,hyperref}
\usepackage{amsmath}
\usepackage{amsfonts}
\usepackage{amssymb}
\usepackage{siunitx}
\usepackage{multirow}
\usepackage{graphicx}
\usepackage{lineno}

\begin{document}
	\nolinenumbers

\begin{frontmatter}

\title{Stability of Neutron Stars with Dark Matter Core Using Three Crustal Types and the Impact on Mass-Radius Relations}

\author[add1]{Adrian G. Abac\corref{cor1}}
\ead{adrian.abac@aei.mpg.de}
\author[add2]{Christopher C. Bernido}
\ead{cbernido.cvif@gmail.com}
\author[add3]{Jose Perico H. Esguerra}
\ead{jesguerra@nip.upd.edu.ph}

\cortext[cor1]{Corresponding author}
\address[add1]{Max Planck Institute for Gravitational Physics (Albert Einstein Institute), Am Mühlenberg 1, Potsdam 14476, Germany}
\address[add2]{Research Center for Theoretical Physics, Central Visayan Institute Foundation, Jagna, Bohol 6308, Philippines}
\address[add3]{Theoretical Physics Group, National Institute of Physics, University of the Philippines, Diliman, Quezon City, Philippines 1101}


\begin{abstract}
	We investigate the effects of dark matter (DM) on the nuclear equation of state (EoS) and neutron star structure, in the relativistic mean field theory, both in the absence and presence of a crust. The $\sigma-\omega$ model is modified by adding a WIMP-DM component, which interacts with nucleonic matter through the Higgs portal. This model agrees well with previous studies which utilized either a more complicated nuclear model or higher-order terms of the Higgs potential, in that DM softens the EoS, resulting in stars with lower maximum masses. However, instabilities corresponding to negative pressure values in the low-energy density regime of the DM-admixed EoS are present, and this effect becomes more prominent as we increase the DM Fermi momentum. We resolve this by confining DM in the star's core. The regions of instability were replaced by three types of crust: first by the Friedman-Pandharipande-Skyrme (FPS), Skyrme-Lyon (SLy) and BSk19 EoS from the Brussels-Montreal Group, which can be represented by analytical approximations. For a fixed value of the DM Fermi momentum $p_F^{DM}$, the DM-admixed neutron star does not have significant changes in its mass with the addition of the crusts. However, the entire mass-radius relation of the neutron star is significantly affected, with an observed increase in the radius of the star corresponding to the mass. The effect of DM is to reduce the mass of the star, while the crust does not affect the radius significantly, as the value of the $p_F^{DM}$ increases.
\end{abstract}


\end{frontmatter}

\nolinenumbers


	\section{\label{Section1}Introduction}
Neutron stars are good testing grounds for predictions of theories beyond the standard model, since they are compact enough to provide conditions necessary for exotic physics to occur \cite{Lattimer2012485,Ozel2016401,Vidana2018,Lavallaz2010}. Furthermore, they are a staple in the studies of nuclear physics, quantum chromodynamics (QCD), and general relativity (GR) \cite{Glendenning2000,Shapiro2004,Camenzind2007}. 

One area of research that is currently very active in theoretical and observational astrophysics are neutron star interiors, especially with the advent of gravitational and electromagnetic wave observations among neutron star mergers \cite{Lattimer2012485,Ozel2016401,Vidana2018,Lavallaz2010}. The description of static, nonrotating neutron stars is achieved by solving the Tolman-Oppenheimer-Volkoff (TOV) equations of GR \cite{Tolman1939,Oppenheimer1939374,Caroll2004}, which are completed by an equation of state (EoS) \cite{Lattimer2012485,Ozel2016401,Haensel2005,Pandharipande1989,Douchin2001,Haensel2007,Potekhin2013}. This yields the mass-radius relations for neutron stars which can be analyzed \cite{Silbar2004892}. Models for neutron stars utilize QCD, or phenomenologically, nuclear field theory in the context of the relativistic mean field theory (rMFT) in obtaining the EoS for nuclear structure, particularly at the core of the star \cite{Chin197424,Serot1997,Dutra2014}. Moreover, several semi-empirical approaches have also been developed to describe the overall structure of the neutron star, by including its outer layers, such as the crust and/or the atmosphere \cite{Haensel2005,Potekhin2013}. 

Another factor that we can consider in the studies of neutron stars are the observations and measurements of the mass-energy density of the universe which shows that majority of its mass-energy content does not come from matter that is well-described by the standard model; about $25\%$ is of the form now known as dark matter (DM) \cite{Calabrese2017,Kisslinger2019}. Strong evidence for the existence of DM using galactic rotation curves was provided by Vera Rubin, Kent Ford and Ken Freeman in the 1960s and 1970s \cite{Rubin1970a,Rubin1970b}. A favored dark matter candidate is the weakly interacting massive particle (WIMP), which is predicted by supersymmetric extensions to the standard model, and at the same time supported by N-body cosmological simulations \cite{Andreas2008,Springer2005}. Reviews on DM can be found in Ref. \cite{Kisslinger2019,Young2017,Arun2017}.

The effects of DM on neutron star structure, and other properties such as tidal deformability, curvature, and inspiral properties of binary neutron stars have been investigated in the literature, using different assumptions on the nature of the DM involved \cite{Goldman1989,Kovaris2010,Panotopoulos2017,Ellis2018,Rezaei2018,Das2019,Kain2021,Das2020,Das2021a,Das2021b}. Some of these used the relativistic mean field theory (rMFT) in quantum hadrodynamics (QHD), starting with different QHD models \cite{Panotopoulos2017,Das2019,Das2020,Das2021a}. In particular, the DM particle is assumed to be fermionic, captured and trapped inside the neutron star \cite{Panotopoulos2017,Das2019,Das2020,Cline2013}. The result of this approach is that DM softens the nuclear equation of state, yielding neutron stars of lower masses than neutron stars without DM \cite{Panotopoulos2017,Das2019,Das2020}. This effect of reducing neutron star masses is also supported by studies assuming that there is a DM core, together with a nuclear EoS in the middle of the star \cite{Ellis2018}.

A nuclear EoS, however is only dominant at the core of the neutron star, with densities greater than $ \rho_c \sim 10^{14}$ g/cm$^3$, while an actual neutron star can have a crust or atmosphere \cite{Haensel2005,Potekhin2013}. The neutron star can then be thought of as having a crust, with density $\rho$, surrounding the core, beginning with density $\rho_c$, such that $\rho <\rho_c$ \cite{Haensel2005,Potekhin2013}. In Ref. \cite{Das2019}, the DM-admixed nuclear EoS was added with a Baym-Pethick-Sutherland (BPS) crust \cite{Baym1971}. The BPS crust however only satisfies the EoS at low densities, and does not include the densities in the crust-core interface, which was approximated in Ref. \cite{Das2019} by a polytropic formula that connects the BPS crust with the DM-admixed core. In this paper we extend these studies by admixing DM at the nuclear core, and by adding three equations of state representing the crust on top of the core: the Friedman-Pandharipande-Skyrme (FPS) EoS, the Skyrme Lyon (SLy) EoS, and the BSk19 EoS, deveopled by the Brussels-Montreal group . 

In this paper, we deal with the simplest QHD model, the $\sigma$-$\omega$ or the Walecka model \cite{Chin197424} and include the Higgs fields $h$ up to order $h^2$. In the Standard Model, the Higgs fields are small fluctuations about the vacuum and higher orders of $h$ can be ignored. The $\sigma$-$\omega$ model describes the interaction between the nucleons in matter through two meson fields, a scalar $\sigma$, and a vector $\omega$, satisfying only the two minimal constraints for nuclear matter: the binding energy per nucleon, and the energy density at saturation. It does not take into account other constraints such as the compression modulus, the effective nucleon mass, isospin symmetry energy, and charge neutrality and beta equilibrium condition \cite{Glendenning2000}. This model also only considers pure neutron matter. Even with the simplicity of the Walecka model, we are still able to extract the implications of putting a crust on top of the core of the star. We then extend the analysis of Ref. \cite{Panotopoulos2017} by investigating instabilities in the DM-admixed EoS, and we fix these instabilities by replacing these unstable regions, which happen to be at the low density-end of the EoS with that of crust EoS, first with the FPS \cite{Pandharipande1989}, then the SLy \cite{Douchin2001}, and then the BSk19 \cite{Goriely2010} EoS; which can be represented by semi-analytical unified models that describe the neutron star crust realistically \cite{Haensel2005,Potekhin2013}. The effects of these modifications to the DM-admixed EoS are then compared and studied.

We summarize the structure of this paper as follows. In Section \ref{Section2}, we discuss the modification of the Walecka model with DM. Section \ref{Section3} then deals with adding the crusts to the DM-admixed EoS. The consequences of these modifications to the neutron star structure are discussed in Section \ref{Section4}. Finally, we conclude by giving some recommendations in Section \ref{Section5}. In this paper, we work with natural units $\hbar = c = 1$ unless otherwise explicitly stated.

\section{\label{Section2}The Walecka Model Equation of State with DM}

The simplest QHD model is the $\sigma-\omega$ or Walecka model \cite{Glendenning2000,Chin197424}. It is a model describing nucleon-nucleon interaction that is mediated by exchanging $\sigma$ and $\omega$ mesons. The fields in this model are based on four particles: the nucleons (neutrons and protons) $\psi$, the scalar meson $\sigma$, and the omega vector mesons $\omega^{\mu}$, with a Lagrangian density given by
\begin{equation}
	\begin{split}
		\mathcal{L}_{\text{had}} =&\bar{\psi} \left[i \gamma_{\mu} \left(\partial^{\mu} + ig_{\omega}\omega^{\mu} \right)-\left(m_n - g_{\sigma}\sigma \right) \right]\psi\\
		&+ \frac{1}{2} \left(\partial_{\mu}\sigma \partial^{\mu}\sigma -m_{\sigma}^2 \sigma^2 \right) - \frac{1}{4} \omega_{\mu \nu}\omega^{\mu \nu} + \frac{1}{2}m_{\omega}^2\omega_{\mu}\omega^{\mu}, \label{eq. 1}
	\end{split}
\end{equation}
where $\omega^{\mu\nu} = \partial^{\mu}\omega^{\nu} - \partial^{\nu}\omega^{\mu}$, $m_n \approx 1$ GeV is the mass of the nucleon (or neutron), $m_{\sigma} = 520$ MeV is the mass of the $\sigma$ meson, $m_{\omega} = 783$ MeV is the mass of the $\omega$ meson, and the dimensionless coupling constants are $g_{\omega}^2 = 190.4$ for the $\omega$ meson coupled to the four-current $\bar{\psi}\gamma^{\mu}\psi$ and $g_{\sigma}^2 = 109.6$ for the $\sigma$ meson coupled with the baryon scalar density $\bar{\psi}\psi$ \cite{Serot1997,Panotopoulos2017}. 

Let us now consider a DM particle with mass $M_{\chi} = 200$ GeV which would be the lightest supersymmetric neutralino \cite{Martin2010}. The fermionic DM Lagrangian density is given by
\begin{equation}
	\begin{split}
		\mathcal{L}_{\text{DM}} =& \bar{\chi}\left[i\gamma^{\mu}\partial_{\mu} - M_{\chi} + yh \right]\chi \\
		&+ \frac{1}{2}\partial_{\mu}h\partial^{\mu}h - \frac{1}{2}M_h^2h^2 + f \frac{m_n}{v}\bar{\psi}h\psi, \label{eq. 2}
	\end{split}
\end{equation}
where we have the Higgs boson $h$ with mass $M_h = 125$ GeV, a DM-Higgs Yukawa coupling $y$, and a nucleon-Higgs Yukawa coupling $fm_n/v$, where $v = 246$ GeV is the Higgs vacuum expectation value, and $f=0.3$ parametrizes the Higgs-nucleon coupling \cite{Martin2010,Murakami2001,Panotopoulos2017,Das2019,Das2020}. Very stringent constraints on the DM-nucleon interaction for DM masses above $6$ GeV are given by recent DM direct detection experiments \cite{Akerib2017,Cui2017,Aprile2018}. We then consider a negligible DM-nucleon coupling and did not include this term in Eq. (\ref{eq. 2}) \cite{Gresham2019,Nelson2019}. The total Langrangian density for the DM-admixed system is then
\begin{equation}
	\mathcal{L} = \mathcal{L}_{\text{had}} + \mathcal{L}_{\text{DM}}. \label{eq. 3}
\end{equation}

In rMFT, the system is assumed to be uniform in its ground state, and the fields in the Lagrangian are replaced by their mean values \cite{Glendenning2000}, that is, $\sigma \rightarrow \langle \sigma \rangle$, $\omega_{\mu} \rightarrow \langle \omega_{\mu}\rangle$, and $h \rightarrow \langle h\rangle$. The equations of motion then become
\begin{equation}
	\begin{split}
		m_{\sigma}^2 \langle \sigma\rangle &= g_{\sigma}\langle \bar{\psi}\psi\rangle\\
		m_{\omega}^2 \langle \omega_{\mu} \rangle &= g_{\omega}\langle \bar{\psi}\gamma_{\mu}\psi\rangle\\
		\left[\gamma_{\mu}(i\partial^{\mu} - g_{\omega}\langle \omega^{\mu}\rangle) - m_n^*\right]\psi(x) &= 0\\
		M_h^2\langle h \rangle &= y \langle \bar{\chi}\chi\rangle + f\frac{m_n}{v}\langle \bar{\psi}\psi\rangle\\
		\left[i\gamma^{\mu}\partial_{\mu} - M_{\chi}^* \right]\chi(x)&=0, \label{eq. 4}
	\end{split}
\end{equation}
where the effective masses are given by
\begin{equation}
	\begin{split} 
		M_{\chi}^* \equiv& M_\chi -y\langle h\rangle\\
		\quad m_n^* \equiv& m_n - g_{\sigma}\langle\sigma\rangle - f\frac{m_n}{v}\langle h\rangle. \label{eq. 5}
	\end{split}
\end{equation}
Defining the following dimensionless quantities to increase the efficiency of our numerical calculations:
\begin{equation}
	\tilde{p} = \frac{p}{m_n}; \quad \varphi = \frac{p_F}{m_n}; \quad \phi = \frac{p_F^{DM}}{m_n}, \label{eq. 6}
\end{equation}
\begin{equation}
	\tilde{\sigma} = \frac{g_{\sigma}\langle\sigma\rangle}{m_n}; \quad \tilde{\omega}_0=\frac{g_{\omega}\langle\omega_0\rangle}{m_n}; \quad \tilde{h} = \frac{Y\langle h\rangle}{m_n}; \quad Y = \frac{fm_n}{v}, \label{eq. 7}
\end{equation}
where $p$ is the particle momentum, $p_F$ is the nucleon Fermi momentum, $p_F^{DM}$ is the DM Fermi momentum, together with the energy density $\epsilon$ and pressure $P$ which forms the parametric EoS,
\begin{equation}
	\tilde{\epsilon} = \frac{\epsilon}{\epsilon_0}; \quad \tilde{P} = \frac{P}{\epsilon_0}; \quad \text{where} \,\, \epsilon_0 = \frac{m_n^4}{3\pi^2}, \label{eq. 8}
\end{equation}
the mean fields for the DM-admixed $\sigma$-$\omega$ model become
\begin{equation}
	\tilde{\sigma} = \frac{g_{\sigma}^2m_n^2}{m_{\sigma}^2\pi^2}\int_0^{\varphi}d\tilde{p}\tilde{p}^2\frac{1-\tilde{\sigma}-\tilde{h}}{\sqrt{\tilde{p}^2 + (1-\tilde{\sigma}-\tilde{h})^2}}, \label{eq. 9}
\end{equation}
\begin{equation}
	\tilde{\omega}_0= \frac{g_{\omega}^2m_n^2}{m_{\omega}^2\pi^2}\frac{\varphi^3}{3}, \label{eq. 10}
\end{equation}
\begin{equation}
	\begin{split}
		\tilde{h} =& \frac{yYm_n^2}{M_h^2\pi^2}\int_0^{\phi}d\tilde{p}\tilde{p}^2\frac{\left(\frac{M_{\chi}}{m_n}-\frac{y}{Y}\tilde{h}\right)}{\sqrt{\tilde{p}^2+\left(\frac{M_{\chi}}{m_n}-\frac{y}{Y}\tilde{h}\right)^2}}\\
		&+ \frac{Y^2m_n^2}{M_h^2\pi^2}\int_0^{\varphi}d\tilde{p}\tilde{p}^2\frac{(1-\tilde{\sigma}-\tilde{h})}{\sqrt{\tilde{p}^2+(1-\tilde{\sigma}-\tilde{h})^2}}. \label{eq. 11}
	\end{split}
\end{equation}
Taking into account only pure neutron matter, our dimensionless, parametric, $\sigma$-$\omega$-DM EoS of the form $\epsilon(P)$ or $P(\epsilon)$ is then written as
\begin{equation}
	\begin{split}
		\tilde{\epsilon} =& \frac{1}{\epsilon_0}\Bigg[\frac{1}{2}\left(\frac{m_{\sigma}m_n}{g_{\sigma}}\right)^2\tilde{\sigma}^2+ \frac{1}{2}\left(\frac{m_{\omega}m_n}{g_{\sigma}}\right)^2\tilde{\omega}_0^2\\
		&+ \frac{1}{2}\left(\frac{M_hm_n}{Y}\right)^2\tilde{h}^2 + \frac{m_n^4}{\pi^2}\int_0^{\varphi}d\tilde{p}\tilde{p}^2 \sqrt{\tilde{p}^2 + (1-\tilde{\sigma}-\tilde{h})^2}\\
		&+ \frac{m_n^4}{\pi^2}\int_0^{\phi}d\tilde{p}\tilde{p}^2 \sqrt{\tilde{p}^2 + \left(\frac{M_{\chi}}{m_n}-\frac{y}{Y}\tilde{h}\right)^2}\Bigg], \label{eq. 12}
	\end{split}
\end{equation}
\begin{equation}
	\begin{split}
		\tilde{P} =& \frac{1}{\epsilon_0}\Bigg[-\frac{1}{2}\left(\frac{m_{\sigma}m_n}{g_{\sigma}}\right)^2\tilde{\sigma}^2 + \frac{1}{2}\left(\frac{m_{\omega}m_n}{g_{\sigma}}\right)^2\tilde{\omega}_0^2\\
		&- \frac{1}{2}\left(\frac{M_hm_n}{Y}\right)^2\tilde{h}^2 + \frac{m_n^4}{3\pi^2}\int_0^{\varphi}d\tilde{p}\frac{\tilde{p}^2}{\sqrt{\tilde{p}^2 + (1-\tilde{\sigma}-\tilde{h})^2}}\\
		&+ \frac{m_n^4}{3\pi^2}\int_0^{\phi}d\tilde{p}\frac{\tilde{p}^4}{\sqrt{\tilde{p}^2 + \left(\frac{M_{\chi}}{m_n}-\frac{y}{Y}\tilde{h}\right)^2}} \Bigg]. \label{eq. 13}
	\end{split}
\end{equation}

To numerically solve the EoS, we first solve simultaneously for the mean fields Eqs. (\ref{eq. 9})-(\ref{eq. 11}), for a range of hardron Fermi momenta $p_F$, and for a given DM Fermi momentum $p_F^{DM}$, before substituting these to the EoS. We take the values of the DM Fermi momenta to be $p_F^{DM} = 0.02 \,\text{GeV}, \, 0.04 \,\text{GeV}, \, 0.06 \,\text{GeV}$, in accordance with existing literature \cite{Das2019}, which also evades constraints from DM search experiments. 

\begin{figure}[htb!]
	\centering
	\includegraphics[scale=0.25]{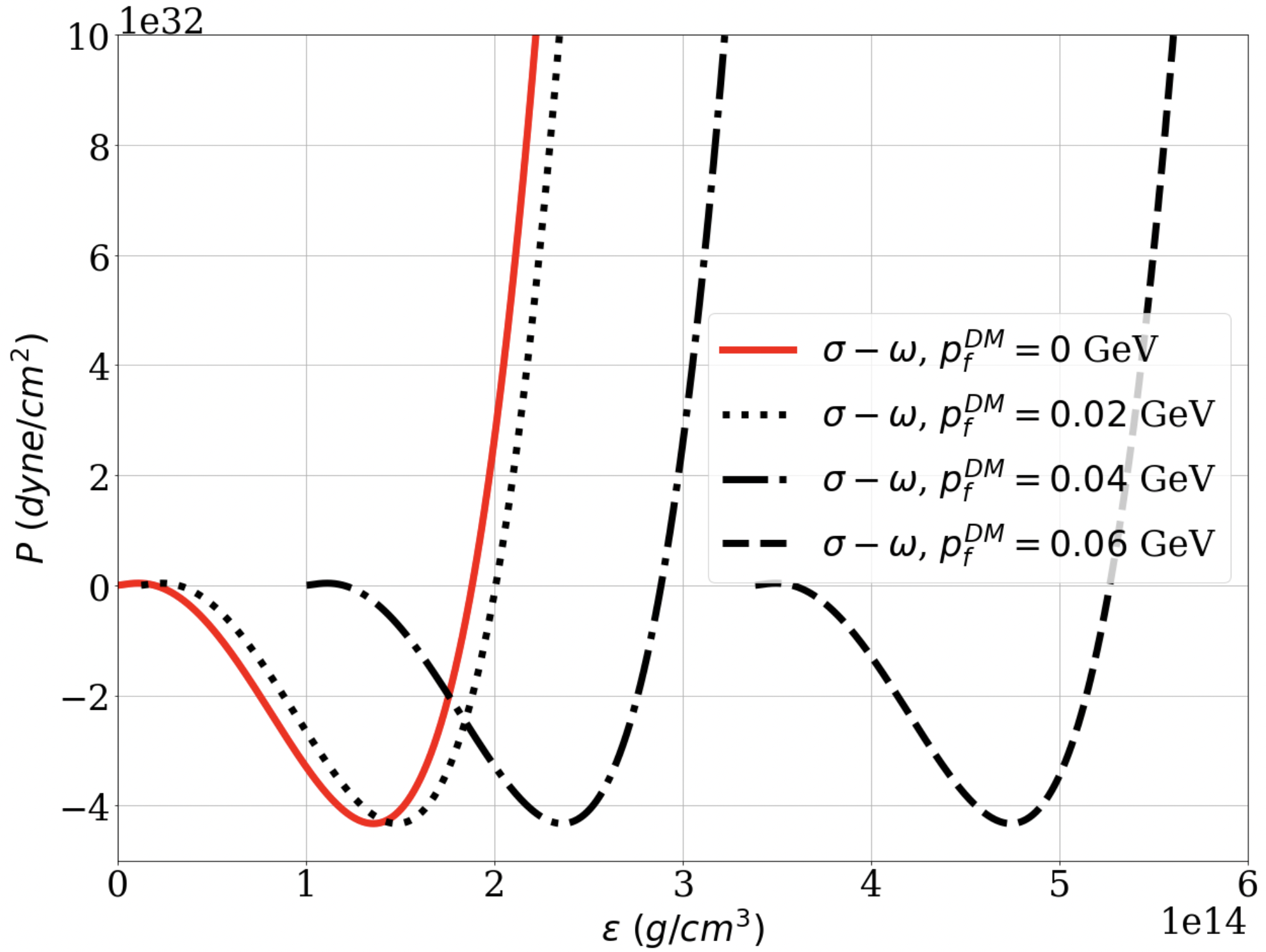}
	\caption{\label{fig1}$\sigma$-$\omega$-DM $P(\epsilon)$ EoS.}
\end{figure}

Figure \ref{fig1} shows the $\sigma$-$\omega$-DM EoS plots for different values of the DM Fermi momentum. The effect of DM indeed is to ``soften" the EoS \cite{Panotopoulos2017}, that is, to shift the EoS towards higher energy density values corresponding to pressure values. This overall trend is also observed in more complicated EoS \cite{Das2019,Das2020}.  The effect becomes more manifest as the value of $p_F^{DM}$ increases. 

We also observe from Fig. \ref{fig1} that the $\sigma$-$\omega$-DM EoS has negative values of pressure in the low-pressure regimes, corresponding to instabilities in the EoS \cite{Schroeder2000}. Moreover, since the EoS only describes neutron matter, this EoS is only applicable at high densities, especially those around greater than $ \rho_c$. To make a more complete description of the star, we can replace the low density regions of the nuclear EoS with a crustal EoS, which also replaces the regions containing negative values of pressure. This ensures the stability of the EoS and corresponding neutron star configuration at low pressures.

\section{\label{Section3}The DM-Admixed EoS with Crust}

A remedy for the instability problem presented in Section \ref{Section2} is to replace the unstable regions in the EoS. This can be done by replacing the low-pressure regions with another EoS that better describes it, similar to an atmosphere or crust. The underlying assumption for this is that our DM is trapped only inside the core of the neutron star; this means that the crust contains a negligible amount of DM particles. We note that this method cannot precisely determine the relative amount of DM inside the core of the star, which would make the model less predictive, as first discussed in Ref. \cite{Das2019}. Nevertheless, we can investigate the implications of having different crusts on top of the DM-admixed neutron core.

The crustal EoS that we consider here are the FPS, SLy, and BSk19 EoS. The FPS and SLy EoS both use effective nucleon-nucleon interactions, but for the FPS, the fitting of ground state properties of laboratory nuclei was not included in the derivation \cite{Pandharipande1989,Douchin2001}. For the SLy EoS, the use of effective NN interactions were combined with the general procedure of fitting the properties of doubly magic nuclei \cite{Douchin2001}; this is suitable for application for the calculation of properties of neutron rich matter. Meanwhile, the EoS dubbed ``BSk19" developed by the Brussels-Montreal group was based on nuclear energy-density functionals, and was derived from generalized Skyrme interactions, supplemented with microscopic contact pairing interactions, a phenomenological Wigner terms, and correction terms for the collective energy \cite{Potekhin2013,Goriely2010}. 

The three EoS can be represented by semi-analytical models, which can describe all the regions of the neutron star interior \cite{Haensel2005,Potekhin2013}. In this study, we use the three EoS to model the crust, as our nuclear core is DM-admixed, which was not considered in Ref. \cite{Haensel2007, Haensel2005,Potekhin2013}. For the FPS and SLy EoS, the parametrization for nonrotating stars \cite{Haensel2005,Potekhin2013} is given by
\begin{equation}
	\begin{split}
		\log P =&\frac{a_1 + a_2 \log \epsilon + a_3 (\log \epsilon)^3}{1 + a_4\log \epsilon}f_0(a_5(\log \epsilon - a_6))\\
		& + (a_7 + a_8\log \epsilon)f_0(a_9(a_{10}- \log \epsilon))\\
		&+(a_{11} + a_{12}\log \epsilon)f_0(a_{13}(a_{14}-\log \epsilon))\\
		&+ (a_{15} + a_{16}\log \epsilon)f_0(a_{17} (a_{18} -\log\epsilon))\\
		&+ \frac{a_{19}}{1 + \left[a_{20}(\log\epsilon - a_{21})\right]^2} + \frac{a_{22}}{1 + [a_{23}(\log\epsilon - a_{24})]^2}, \label{eq. 16}
	\end{split}
\end{equation}
where the units for $P$ and $\epsilon$ are in $\text{dyne/cm}^2$ and $\text{g/cm}^3$, respectively, the $a_i \,\, (i = 1, 2, 3,..., 24)$ are fitting constants, and the function $f_0(x)$ is defined as
\begin{equation}
	f_0(x) = \frac{1}{e^x +1}. \label{eq. 17}
\end{equation}

For the BSk19 EoS, additional terms were present in the analytical approximation that improve the fit near the boundaries between the outer and inner crust, and between the crust and core \cite{Potekhin2013} (in the SLy EoS, the changes were less abrupt \cite{Haensel2005}). The values of $a_i$, taken from Ref. \cite{Haensel2005,Potekhin2013}, are given in Table \ref{Table1}.
\begin{table}[hbt!]
\centering
	\caption{\label{Table1}Fitting Parameters for the FPS, SLy, and BSk19 EoS \cite{Haensel2005,Potekhin2013}}
	\begin{tabular}{c|c|c|c}	
	i & $a_i$, FPS & $a_i$, SLy & $a_i$, BSk19 \\
	\hline
	1 & 6.22 & 6.22 & 3.916\\
	2 & 6.121 & 6.121 & 7.701\\
	3 & 0.006004 & 0.005925 & 0.00858\\
	4 & 0.16345 & 0.16326 & 0.22114\\
	5 & 6.50 & 6.48 & 3.269\\
	6 & 11.8440 & 11.4971  & 11.964\\
	7 & 17.24 & 19.105 & 13.349\\
	8 & 1.065 & 0.8938 & 1.3683\\
	9 & 6.54 & 6.54  & 3.254\\
	10 & 11.8421 & 11.4950 & 11.964 \\
	11 & -22.003 & -22.775 & -12.953 \\
	12 & 1.5552 & 1.5707 & 0.9237\\
	13 & 9.3 & 4.3 & 6.20\\
	14 & 14.9 & 14.08 & 14.383\\
	15 & 23.73 & 27.80 & 16.693\\
	16 & -1.508 & -1.653 & -1.05146\\
	17 & 1.79 & 1.50 & 2.48 \\
	18 & 15.13 & 14.67 & 15.362\\
	19 & 0 & 0 & 0.085\\
	20 & 0 & 0 & 6.238\\
	21 & 0 & 0 & 11.6\\
	22 & 0 & 0 & -0.029\\
	23 & 0 & 0 & 20.1\\
	24 & 0 & 0 &  14.19\\
\end{tabular}

\end{table}
		
\begin{figure}[htb!]
	\centering
	\includegraphics[scale=0.3]{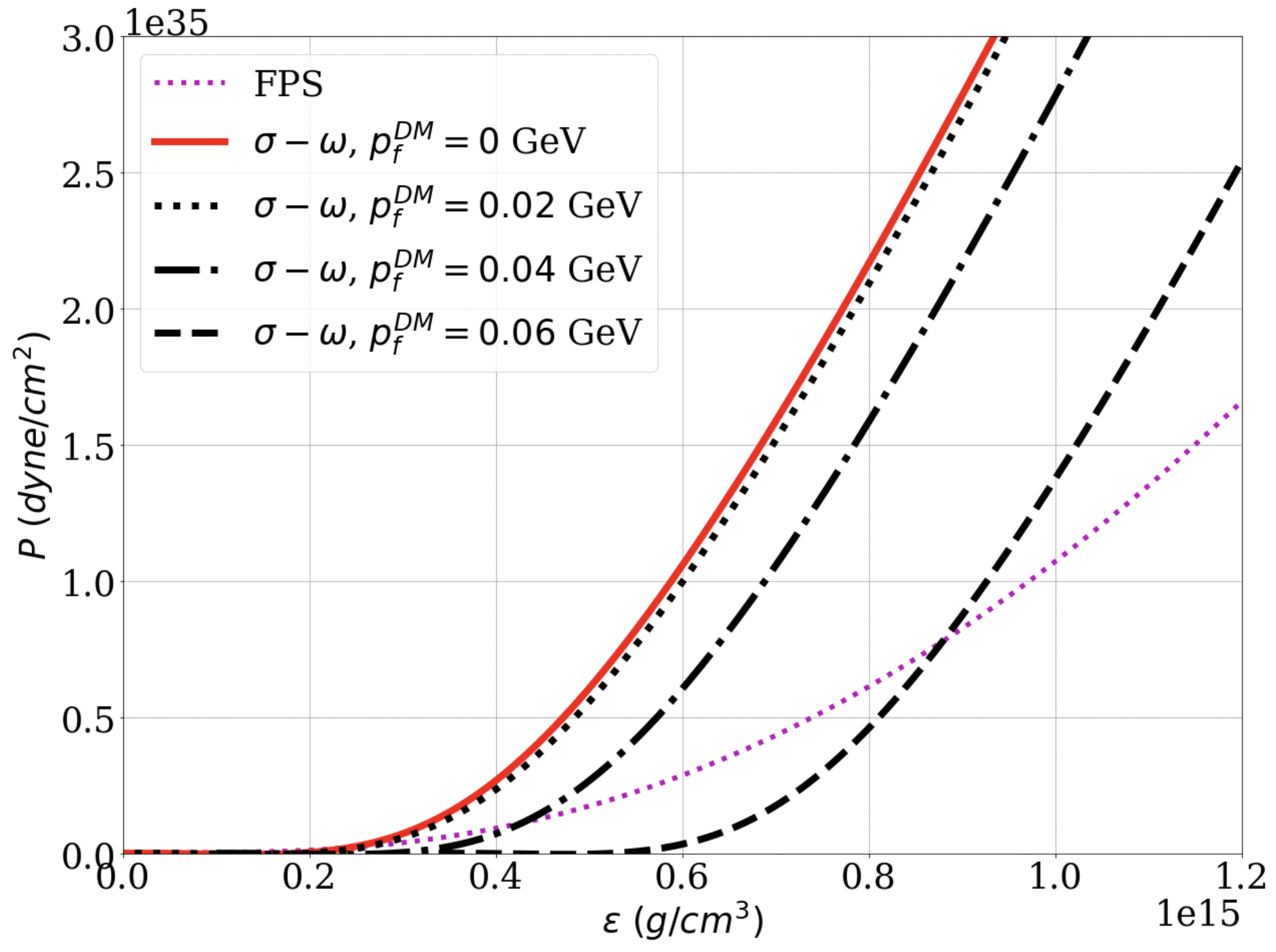}
	\caption{\label{fig2}DM-admixed EoS with the crust (FPS) EoS}
\end{figure}

We plot the FPS EoS superimposed on the DM-admixed EoS, and the result is shown in Fig. \ref{fig2}. To fix the instabilities in the DM-admixed nuclear EoS, we replace the unstable regions of the EoS with the crust EoS right up to the point of intersection, to ensure the continuity in the pressure of the EoS. Because of the effect of DM on the $\sigma$-$\omega$ EoS, the point of intersection between the crust and the nuclear EoS also occurs at higher values of the energy densities. This is also the case for the SLy and BSk19 EoS. These points of intersection are found in Table \ref{Table2}, where $\epsilon$ and $P$ are in ($\times 10^{14}$ g/cm$^3$) and ($\times 10^{34}$ dyne/cm$^2$), respectively. From the intersection, the pressure then decreases towards the edge of the star, while it increases towards the star's center. 

\begin{table}[htb!]
	\centering
	\caption{\label{Table2} Energy Densities and Pressures at Intersection Between the $\sigma$-$\omega$-DM and Crust EoS. The unit for $\epsilon$ is ($\times 10^{14}$ g/cm$^3$), while $P$ is in ($\times 10^{34}$ dyne/cm$^2$).}
		\begin{tabular}{c|cc|cc|cc}	
			
			& \multicolumn{2}{c|}{\bf{FPS-$\sigma$-$\omega$-DM}} & \multicolumn{2}{c|}{\bf{SLy-$\sigma$-$\omega$-DM}} &\multicolumn{2}{c}{\bf{BSk19-$\sigma$-$\omega$-DM}} \\
			$p_F^{DM}$ (GeV) & $\epsilon$  & $P$  & $\epsilon$ & $P$ & $\epsilon$ & $P$\\
			\hline
			0 & 2.42 & 0.199 & 2.58 & 0.303 & 2.45 & 0.217\\
			0.02 & 2.65 & 0.262 & 2.81 & 0.388 & 2.65 & 0.268 \\
			0.04 & 4.28 & 1.11 & 4.55 & 1.60 & 4.19 & 0.976\\
			0.06 & 8.79 & 7.78 & 10.2 & 14.8 & 8.69 & 7.34 	 
		\end{tabular}
\end{table}
\begin{figure}[htb!]
	\centering
	\includegraphics[scale=0.43]{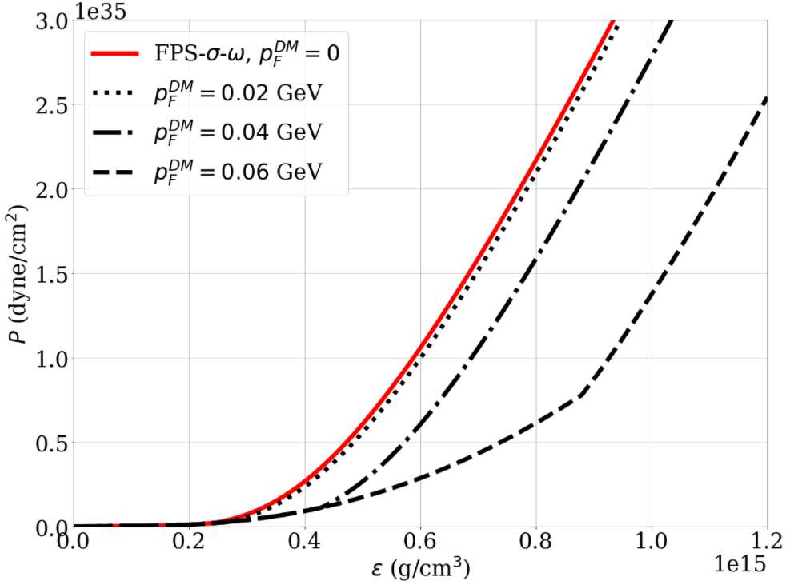}
	\caption{\label{fig3}FPS-$\sigma$-$\omega$-DM EoS}
\end{figure} 

\begin{figure}[htb!]
	\centering
	\includegraphics[scale=0.43]{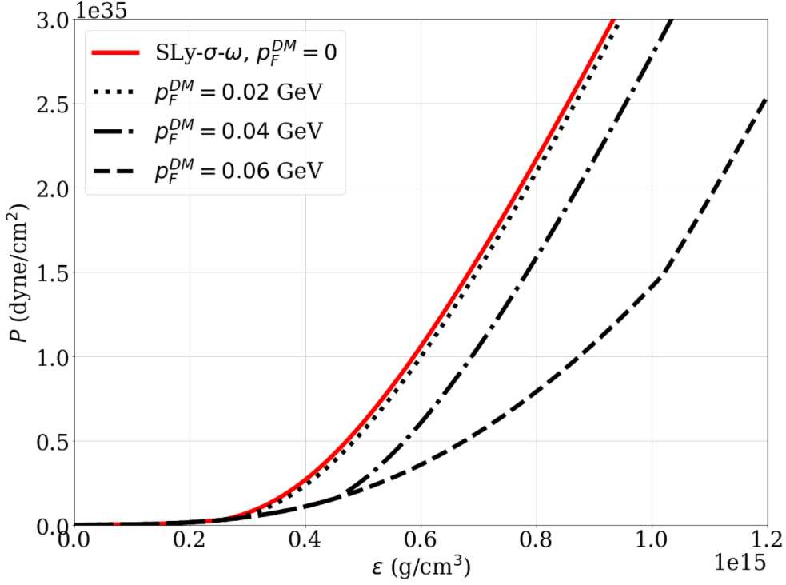}
	\caption{\label{fig4}SLy-$\sigma$-$\omega$-DM EoS}
\end{figure}

\begin{figure}[htb!]
	\centering
	\includegraphics[scale=0.43]{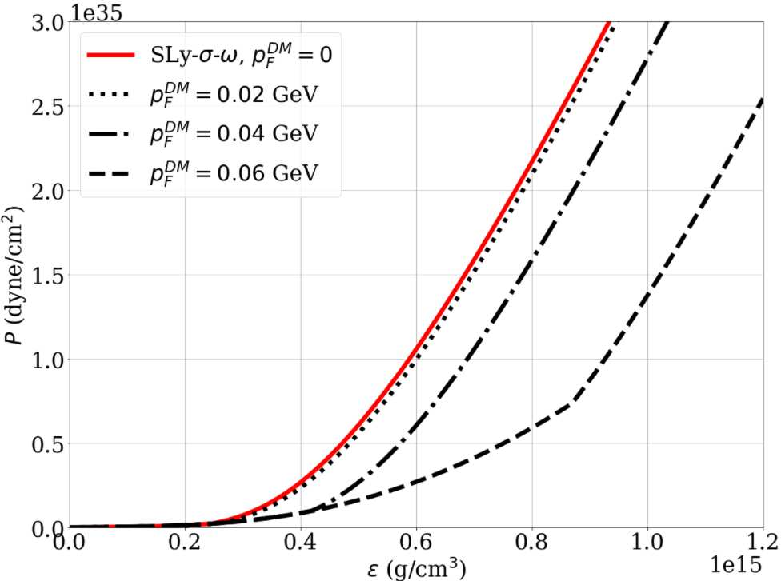}
	\caption{\label{fig5}BSk19-$\sigma$-$\omega$-DM EoS}
\end{figure}

The resulting EoS with the FPS, SLy, and BSk-19 EoS are found in Figs. \ref{fig3}-\ref{fig5}. We see that the the overall EoS has significant changes in the crust-core interface as well as the crust as low densities. The SLy EoS is observed to intersect with the DM-admixed EoS at higher energy densties, and correspondingly, higher pressures, followed by the FPS and BSk19 EoS. We note the similarity between the EoS modified by FPS and BSk19, especially at their points of intersection, but it will be later shown that these EoS result to different mass-radius relations. The DM effects also only occur after the intersection points, since they are admixed in the nuclear EoS. 

In summary, the stability of our neutron star EoS is achieved by removing the negative pressure regions of the EoS through the replacement of these regions by the EoS of the crusts, surrounding the core where DM is confined. This also ensures that the model is stable even with higher values of the DM Fermi momentum. It will be interesting now to see how these changes in the EoS brought by the addition of the crustal EoS, affect the mass-radius relations of neutron stars, and we investigate this in the next section.

\section{\label{Section4}The Structure Equations and Mass-Radius Relations}
Using the dimensionless quantities for $\epsilon$ and $P$ defined in Eq. (8) as well as the following:
\begin{equation}
	\tilde{M} = \frac{M}{M_{\odot}}, \quad \tilde{r} = \frac{r}{R_0}, \quad R_0 = GM_{\odot}, \quad \Omega = \frac{4\pi\epsilon_0}{M_{\odot}}R_0^3,
\end{equation}
where $M$ is the mass, $r$ is the distance from the center of the star, and $M_{\odot}$ is the solar mass, we can write the TOV equations in dimensionless form as
\begin{equation}
	\frac{d\tilde{P}}{d\tilde{r}} = \frac{-\left[\tilde{\epsilon}+\tilde{P}
		\right]\left[\tilde{M}+\Omega\tilde{r}^3\tilde{P}\right]}{\tilde{r}^2 - 2\tilde{M}\tilde{r}},
\end{equation}
\begin{equation}
	\frac{d\tilde{M}}{d\tilde{r}} = \Omega\tilde{r}^2\tilde{\epsilon}, 
\end{equation}
with the conditions
\begin{equation}
	\begin{split}
		\tilde{P}(\tilde{r}=0)=\tilde{P}_c, \quad \tilde{P}(r=R_{\star}) = 0\\
		\tilde{M}(\tilde{r}=0)=0, \quad \tilde{M}(r=R_{\star}) = M_{\star},
	\end{split}
\end{equation}
where $P_c$ is the central pressure and $R_{\star}$ is the stellar radius. The TOV equations describe static, spherically symmetric, nonrotating stars in GR \cite{Glendenning2000,Camenzind2007,Caroll2004}. The modified EoS from Section \ref{Section3} are fed into the TOV equations and solved numerically using the forward Euler method over a range of central pressures $\tilde{P}_c$. The initial condition is such that the pressure is greatest at the center of the star, and reaches zero at the star's edge, defining the stellar radius at $r=R_{\star}$. Meanwhile, the equation for the mass is cumulative, such that it reaches the stellar mass $M = M_{\star}$ at $R_{\star}$. For a range of central pressures, we can then form a parametric relation between $M_{\star}$ and $R_{\star}$, known as the mass-radius relation of the star \cite{Glendenning2000}. We now investigate in this section the effects of three modified EoS that we obtained in Section \ref{Section3} to the mass-radius relations of neutron stars. 
\begin{figure}[htb!]
	\centering
	\includegraphics[scale=0.43]{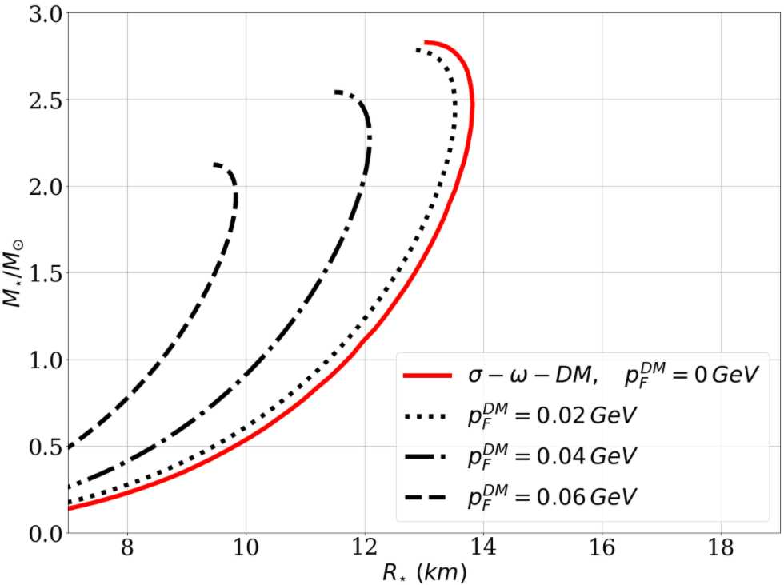}
	\caption{\label{fig6}Mass-Radius Relation, $\sigma$-$\omega$-DM EoS}
\end{figure}

\begin{figure}[htb!]
	\centering
	\includegraphics[scale=0.43]{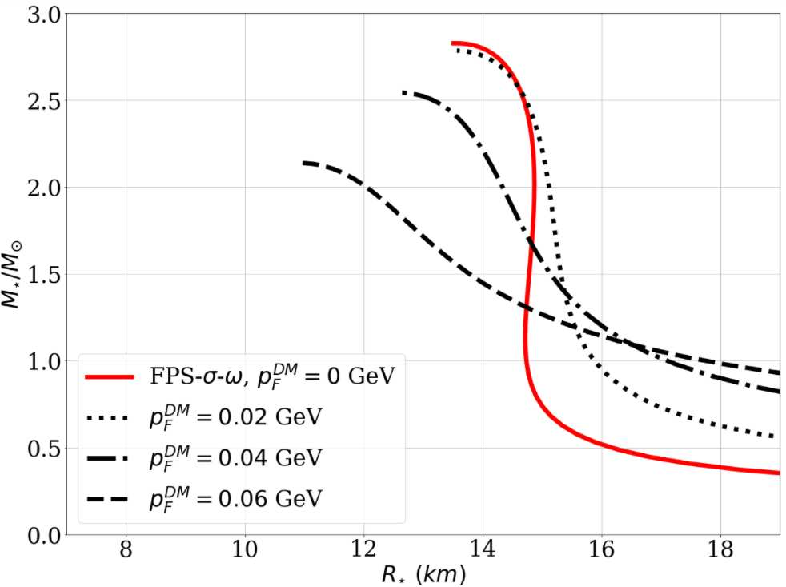}
	\caption{\label{fig7} Mass-Radius Relation, FPS-$\sigma$-$\omega$-DM EoS}
\end{figure}

\begin{figure}[htb!]
	\centering
	\includegraphics[scale=0.43]{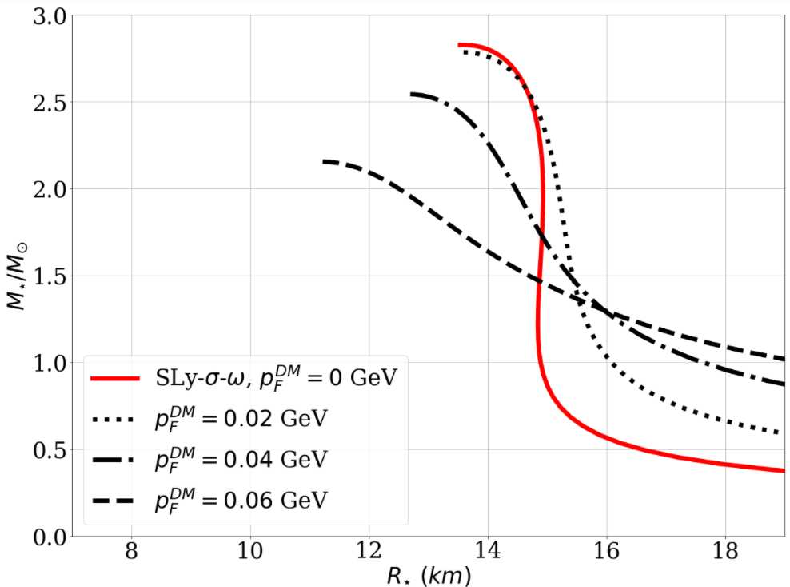}
	\caption{\label{fig8}Mass-Radius Relation, SLy-$\sigma$-$\omega$-DM EoS}
\end{figure}

\begin{figure}[htb!]
	\centering
	\includegraphics[scale=0.43]{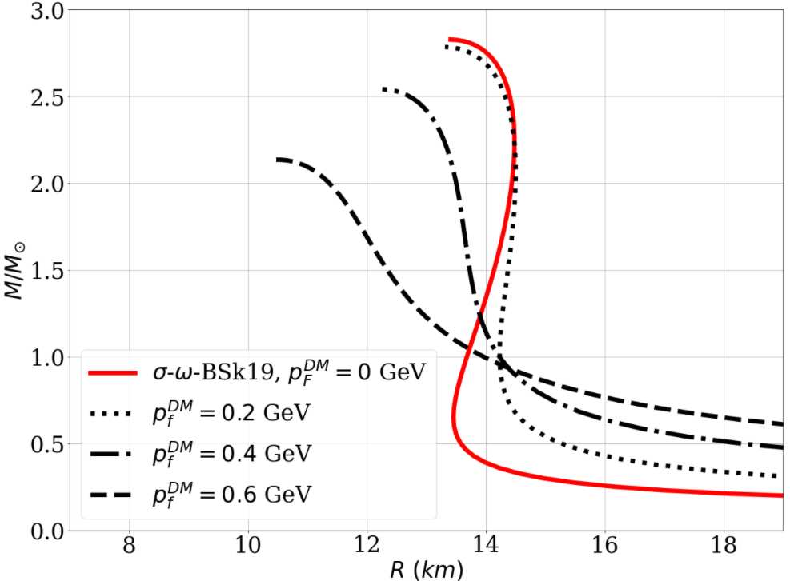}
	\caption{\label{fig9}Mass-Radius Relation, BSk19-$\sigma$-$\omega$-DM EoS}
\end{figure}

As a reference, we also obtain the mass-radius relations for the $\sigma$-$\omega$-DM model for different values of $p_F^{DM}$, without any crust, which is similar to the results of Ref. \cite{Panotopoulos2017}. These are shown in Figure \ref{fig6}. Meanwhile, the mass-radius relations for the FPS-$\sigma$-$\omega$-DM EoS, SLy-$\sigma$-$\omega$-DM EoS, and BSK19-$\sigma$-$\omega$-DM EoS are shown in Figs. \ref{fig7}-\ref{fig9}. Note that the addition of the crust produces significant changes to the mass-radius relations, by increasing the radii of the neutron star corresponding to the mass. However, the effect of DM remains generally the same: to ``shrink" the neutron star, by producing stars of lower masses and smaller radii as the value of $p_F^{DM}$ gets larger (see Table 3). We also note that the limiting/maximum mass of the neutron star increases by small amounts for the EoS. 

We also observe from Figs. \ref{fig7}-\ref{fig9} that for every nonzero $p_F^{DM}$, each plot of the mass-radius relations intersects that of the original $\sigma$-$\omega$-DM EoS ($p_F^{DM} = 0$). The stellar masses on the right side of this intersection are increased from that without DM as a function of increasing $p_F^{DM}$, while the masses decrease in the left of the intersection point as a function of increasing $p_F^{DM}$. We also note that in the absence of DM, that is, at $p_F^{DM} = 0$, the mass-radius relation for all crusted neutron stars yield more or less the same maximum masses (see Table 3).

The FPS- and SLy- crusted mass-radius relations are comparable due to their similarities in their approach of modelling neutron star interiors \cite{Haensel2005}. The mass-radius relations for the FPS- and SLy- crust stars intersect at around $M_{\star} \approx 1.3M_{\odot}$. In Ref. \cite{Haensel2005}, the main differences between the two EoS occur at the crust-core interface ($\rho \approx 10^{12}-10^{15}$ g/cm$^3$). Meanwhile, the intersection of the mass-radius relations for nonzero $p_F^{DM}$ in the BSk19-crusted star occur at around $M_{\star} \approx 0.9M_{\odot}$. The changes to the mass-radius relations are also significant for large values of $p_F^{DM}$. 

From the mass-radius relations, we can then obtain the maximum or limiting masses and corresponding limiting radii for the neutron star for the different EoS, and also for varying values of $p_F^{DM}$. The results are summarized in Table \ref{Table3}. 
\begin{table}
	\centering 
	\caption{\label{Table3} Maximum Masses and Limiting Radii among various EoS}
		\hspace*{-2cm}
		\begin{tabular}{c|cc|cc|cc|cc}
			& \multicolumn{2}{c|}{\bf{$\sigma$-$\omega$-DM}} & \multicolumn{2}{c|}{\bf{FPS-$\sigma$-$\omega$-DM}} & \multicolumn{2}{c|}{\bf{SLy-$\sigma$-$\omega$-DM}} &\multicolumn{2}{c}{\bf{BSk19-$\sigma$-$\omega$-DM}} \\
			$p_F^{DM}$ (GeV) & $M_{\text{lim}}/M_{\odot}$ & $R_{\text{lim}}$ (km) & $M_{\text{lim}}/M_{\odot}$ & $R_{\text{lim}}$ (km) & $M_{\text{lim}}/M_{\odot}$ & $R_{\text{lim}}$ (km) & $M_{\text{lim}}/M_{\odot}$ & $R_{\text{lim}}$ (km)\\
			\hline
			0 & 2.827 & 13.039 & 2.827 & 13.512 & 2.827 & 13.533 & 2.827 & 13.406\\
			0.02 & 2.785 & 12.836 & 2.785 & 13.567 & 2.785 & 13.593 & 2.784 & 13.311\\
			0.04 & 2.541 & 11.490 & 2.542 & 12.532 & 2.543 & 12.657 & 2.541 & 12.257\\
			0.06 & 2.124 & 9.447 & 2.138 & 10.899 & 2.155 & 11.220 & 2.135 & 10.461	 
		\end{tabular}
	\hspace*{-2cm}
\end{table}

\begin{figure}[htb!]
	\centering
	\includegraphics[scale=0.42]{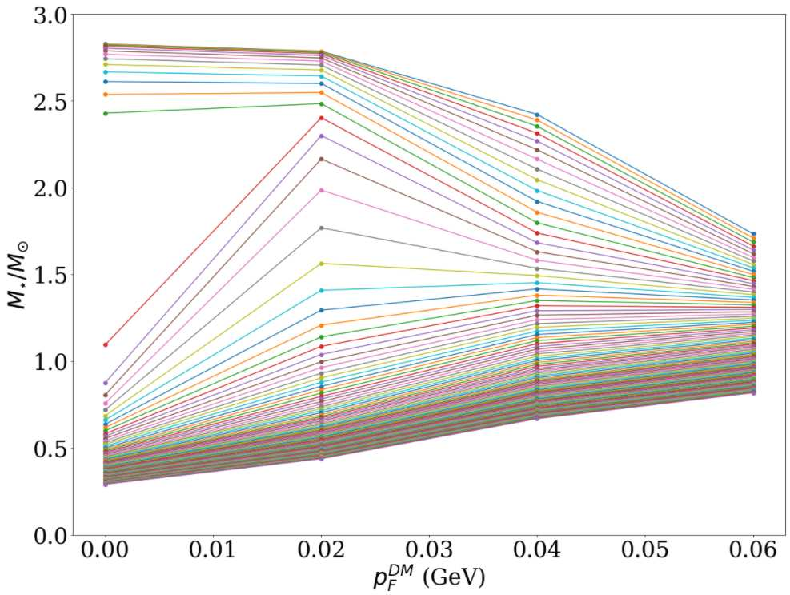}
	\caption{\label{fig10}$M_{\star}/M_{\odot}$ as a function of $p_F^{DM}$, SLy-$\sigma$-$\omega$-DM EoS}
\end{figure}

To further investigate the aforementioned intersections between the mass-radius relations of the crusted stars with nonzero $p_F^{DM}$, we can plot the stellar mass $M_{\star}$ as a function of the DM Fermi momentum $p_F^{DM}$, for some constant stellar radius $R_{\star}$. The result for the SLy-crusted star is shown in Figure \ref{fig10}. Each line in Figure \ref{fig10} corresponds to one fixed radius $R_{\star}$ corresponding to different masses $M_{\star}/M_{\odot}$ as the $p_F^{DM}$ increases. The radii are separated by an interval of $0.1$ km, and each line changes shape for every value of $R_{\star}$. Lines that start from the left at $M_{\star}/M_{\odot} \lesssim 1.0$ correspond to different radii greater than $16$ km. Lines that start at $M_{\star}/M_{\odot} \gtrsim 1.5$ correspond to different radii lower than $15$ km. The different lines of constant radius tend to approach a value of $M_{\text{con}}/M_{\odot} \simeq 1.3$, as $p_F^{DM}$ increases, and this mass corresponds to radius $R_{\star} \simeq 15 \,\, \text{to} \,\, 16$ km, and central pressure $\tilde{P}_c \simeq 0.01 \,\, \text{to} \,\, 0.03$. From this, we can speculate that a DM-admixed compact object, which may not necessarily be a neutron star, could potentially exist, with size (mass and radius) that is conducive to a wide range of values of the DM Fermi momentum $p_F^{DM}$. We emphasize that the DM in the star, as previously mentioned, does not interact with the nucleons, and only interacts with the Higgs particle. This star may be comprised mostly of DM and Higgs particles but not yet detectable by current observational means.

The same behavior is observed for the FPS- and BSk19-crusted star, albeit with slightly different values of $M_{\text{con}}$ and corresponding $R_{\star}$. For the BSk19-crusted star, the masses seemingly converge at $M_{\text{con}}/M_{\odot} \lesssim 0.9$ with increasing $p_F^{DM}$. This mass is below that of typical neutron stars but may possibly indicate a compact object which accommodates a wide range of $p_F^{DM}$ whose radius is around $14.5$ km.

\section{\label{Section5}Conclusions}
In this work, we extended the investigation of DM-admixed neutron stars by confining the DM in the star's core and by adding a crust on the core. This simulates a neutron star with a crust, and with a core dominated by the nuclear equation of state, which, in this case, was the $\sigma$-$\omega$ model added with DM, which was obtained via relativistic mean field theory. 

Three types of crust were considered: the FPS, SLy, and BSk19 crusts. These crust equations of state were used to replace the instabilities in the $\sigma$-$\omega$-DM EoS, corresponding to negative values of the pressure at the lower density regime. This is interpreted as confining the DM to the core of the neutron star, with the crust surrounding it. This stability of the EoS is also achieved even at higher values of the DM Fermi momentum. The resulting mass-radius relations are markedly different from neutron stars without crust \cite{Panotopoulos2017}. DM effects are primarily responsible for decreasing the star mass, while the main effect of the crust is to increase the star radius. We also note that, with or without the crust, both the maximum mass and limiting radius of the neutron star progressively decreases as the DM Fermi momentum $p_F^{DM}$ increases in value (see Table \ref{Table3}). We also speculate the possibility of a compact object, containing a DM core and a crust, existing with a mass and radius that accommodate a wide range of values for $p_F^{DM}$ (see Figure \ref{fig10}).

One can then extend this study by using more complicated models for the nuclear equation of state, to address the limitations of the $\sigma$-$\omega$ model. To account for more realistic physics inside neutron stars, one has to consider other constraints for nuclear matter beyond the two minimal constraints satisfied by the $\sigma$-$\omega$ model. The constraints include the compression modulus and effective nucleon mass at saturation, which are resolved by adding the self interaction terms $\sim \sigma^3$ and $\sim \sigma^4$, the isospin symmetry energy which is fixed by adding the $\rho$ meson, and the charge neutrality and beta-equilibrium conditions which are resolved by adding leptons. This will be the second part of this study. The effects of DM may also be investigated on the star's crust, taking into account the relative amounts of DM that must be present in either the core or the crust. Observations of neutron stars and neutron star mergers may also give constraints on the parameters of DM and the models used in this study.

\section*{\label{Acknowledgments}Acknowledgments}
A. G. Abac wishes to acknowledge the support of the Department of Science of Technology (Philippines) - Accelerated Science and Technology Human Resource Development Program during the course of this study.

\bibliographystyle{elsarticle-num-names}
\bibliography{Abac_DM_NS}

\end{document}